\documentclass[conference]{IEEEtran}

\usepackage{graphicx}
\usepackage{subcaption}
\usepackage{balance}
\usepackage{comment}
\usepackage{amsmath}
\usepackage[bookmarks=false]{hyperref}
\usepackage[numbers]{natbib}
\usepackage{fancyhdr}
\usepackage{lipsum} 
\ifCLASSINFOpdf
\else
\fi

\usepackage{balance}

\usepackage{multirow}
\usepackage[table]{xcolor}
\fancypagestyle{accepted}{
    \fancyhf{}
    
    \fancyfoot[L]{\footnotesize \copyright 2019 IEEE. Personal use of this material is permitted. Permission from IEEE must be obtained for all other uses, in any current or future media, including reprinting/republishing this material for advertising or promotional purposes, creating new collective works, for resale or redistribution to servers or lists, or reuse of any copyrighted component of this work in other works. DOI: \href{https://doi.org/10.1109/IGSC48788.2019.8957182}{10.1109/IGSC48788.2019.8957182}}
}

\begin{document}
	\title{SCART: Predicting STT-RAM Cache Retention Times Using Machine Learning 
}

	\author{\IEEEauthorblockN{Dhruv Gajaria, Kyle Kuan, and
	Tosiron Adegbija}
	\IEEEauthorblockA{Department of Electrical \& Computer Engineering \\ University of Arizona, Tucson, AZ, USA \\
	Email: \{dhruvgajaria, ckkuan, tosiron\}@email.arizona.edu}}

	
\maketitle
\thispagestyle{accepted}


\begin{abstract}
Prior studies have shown that the retention time of the non-volatile spin-transfer torque RAM (STT-RAM) can be relaxed in order to reduce STT-RAM's write energy and latency. However, since different applications may require different retention times, STT-RAM retention times must be critically explored to satisfy various applications' needs. This process can be challenging due to exploration overhead, and exacerbated by the fact that STT-RAM caches are emerging and are not readily available for design time exploration. This paper explores using known and easily obtainable statistics (e.g., SRAM statistics) to predict the appropriate STT-RAM retention times, in order to minimize exploration overhead. We propose an STT-RAM Cache Retention Time (SCART) model, which utilizes machine learning to enable design time or runtime prediction of right-provisioned STT-RAM retention times for latency or energy optimization. Experimental results show that, on average, SCART can reduce the latency and energy by 20.34\% and 29.12\%, respectively, compared to a homogeneous retention time while reducing the exploration overheads by 52.58\% compared to prior work.
\end{abstract}


\IEEEpubidadjcol
\section{Introduction}
Spin-transfer torque RAM (STT-RAM) has emerged as a popular alternative to SRAM for implementing caches. STT-RAMs offer several benefits, such as high density, low leakage power, compatibility with CMOS, high endurance, etc. However, STT-RAMs suffer from high write latency and write energy, which may impede their widespread adoption in state-of-the-art resource-constrained systems. A promising optimization involves relaxing STT-RAM's \textit{retention time}---the duration for which data is retained in the absence of power---from the intrinsic duration, which could be up to 10 years \cite{Smullen11}. Reducing the retention time offers much promise for latency and energy improvements because the long write latency and high write dynamic energy directly result from the long retention times of a non-volatile STT-RAM \cite{Smullen11}. Thus, prior works \cite{LARS8342053}, \cite{Smullen11}, \cite{Sun11}, \cite{Jog12} have studied the benefits of reducing/relaxing the retention times, especially in caches since cache data blocks are usually only needed in the cache for short periods of time (typically less than 1 second).

Given a relaxed retention STT-RAM cache (hereafter referred to simply as STT-RAM cache), prior work has shown that different applications may require different retention times. An application's retention time requirements are dictated by its \textit{cache block lifetimes}, i.e., how long the blocks must remain in the cache. To yield maximal benefits from STT-RAM caches, the retention time must be specialized to the needs of the executing applications or application domains. If the retention times are not specialized, they may be over-provisioned, thus wasting energy/latency, or under-provisioned, thus requiring additional schemes (e.g., the dynamic refresh scheme \cite{Sun11}) to maintain data integrity after the retention time elapses. Both cases accrue overheads that may substantially limit optimization potential \cite{Jog12,LARS8342053}. 


To enable right-provisioned retention times for STT-RAM caches, the retention times must be critically explored for different applications and metrics (e.g., energy, latency). An exhaustive exploration of retention times is a challenging task, given that a wide variety of applications, application characteristics (e.g., read/write behaviors, cache block characteristics), and objective functions (e.g., energy, latency, energy delay product, user experience) must be considered. Furthermore, in systems with adaptable retention times, such as the logically adaptable retention STT-RAM (LARS) cache proposed in \cite{LARS8342053}, an exhaustive exploration can incur substantial runtime overheads, including hardware, switching, time, and energy, especially in complex systems. 

\IEEEpubidadjcol
In this paper, we propose an approach---\textit{\textbf{S}TT-RAM \textbf{Ca}che \textbf{R}etention \textbf{T}ime (SCART) Model}---that utilizes machine learning to predict right-provisioned retention times for a variety of systems, applications, and metrics. Since SRAM caches are widely available and accessible to researchers and designers, whereas STT-RAM caches are still nascent, we explore using SRAM characteristics that can easily be obtained via simulations as input labels to enable the prediction of right-provisioned retention times for STT-RAM caches for target applications or application domains. During runtime in a system with multiple retention time units (e.g., \cite{LARS8342053}), based on execution statistics from one cache unit (SRAM, in a hybrid design \cite{Hybrid_cache2} or STT-RAM), our approach can directly predict the best unit on which to run the application, without the need for overhead-prone design space exploration. 

Our contributions are summarized as follows:

\begin{itemize}
    \item We show, for the first time (to our knowledge), that right-provisioned retention times for STT-RAM caches can be predicted using easily obtainable SRAM characteristics.
    \item We compare several machine learning classifiers, and propose a machine learning-based model (\textit{SCART}) that enables fast runtime retention time prediction. SCART can be implemented with low overhead for runtime prediction in a system with multiple retention times or in a hybrid system. 
    \item Using extensive simulations with three benchmark suites (SPEC CPU2006 \cite{spec2006}, MiBench \cite{mibench}, and GAP \cite{beamer2015gap}), to represent different kinds of applications, we show that our model reduces exploration time by 52.58\%. Furthermore, in a runtime implementation, our approach achieves average latency and energy savings of 20.34\% and  29.12\%, respectively, compared to a homogeneous system. 
    
\end{itemize}

\section{Related Work}
The STT-RAM bit cell's basic structure comprises of a transistor and a magnetic tunnel junction (MTJ). STT-RAM's characteristics and operations of the STT-RAM have been discussed in the prior work \cite{Chun13}. Smullen et al. \cite{Smullen11} showed that for implementation in caches, STT-RAM's retention time can be substantially reduced (e.g., by reducing the planar area) in order to mitigate the attendant write latency and energy overheads of non-volatile STT-RAMs. In this section, we summarize a few related prior works that leverage reduced retention STT-RAMs and briefly overview prior work on cross-architectural prediction to motivate our work.

\subsection{Multi-retention and Hybrid STT-RAM Caches}

Sun et.al. \cite{Sun11} proposed to use a  hybrid STT-RAM L2 cache with multiple retention times in order to more closely match the needs of executing applications. The authors used a coarse-grained approach, featuring a long retention time for read-intensive applications and a short retention time for write-intensive applications. Cache blocks that needed to remain in the cache beyond the retention time were refreshed via a DRAM-style dynamic refresh scheme to maintain data correctness. 
To reduce the overheads introduced by the need to refresh cache blocks, Kuan et.al \cite{LARS8342053} further analyzed application cache block characteristics and showed that the refresh overheads could be mitigated by more closely matching the applications' runtime execution requirements. The authors proposed a logically adaptable retention STT-RAM (LARS) L1 cache featuring multiple retention time units, and used a sampling-based algorithm to dynamically determine applications' right-provisioned retention times. 

Since STT-RAM is generally more prone to overheads when running write-intensive applications, due to the high write latency, hybrid (SRAM+STT-RAM) caches have been proposed. To minimize overheads, the STT-RAM is used to run read-intensive workloads and the SRAM is used for write-intensive workloads. While multiple hybrid (SRAM+STT-RAM) caches \cite{Hybrid_cache2} have been proposed, they typically only feature a single retention time. We anticipate that hybrid caches featuring multiple retention times will be explored in the near future. In all these systems, an important existing challenge, which our work addresses, is how to rapidly explore the right-provisioned retention times with which to design the systems, or how to rapidly select the best retention time during runtime, in order to maximize the energy or latency benefits of reduced retention STT-RAM caches. 

\subsection{Cross-Architectural Prediction}
The work proposed herein is along the lines of prior work where a known architecture is used to predict the behavior of an unknown architecture. For instance, Ardalani et.al. \cite{cross_arch} presented cross-architecture performance prediction using CPU implementation to predict the performance of GPUs. Yang et.al. \cite{cross_plt} presented techniques for predicting the performance of parallel applications using partial execution.
Guo et.al. \cite{GPU-GPU} presented a model to provide inter-architecture performance prediction for sparse matrix vector multiplication to help researchers choose the appropriate GPU architecture for the application. Similarly, Zheng et.al. \cite{phase-level-predictor} presented a phase level cross-platform prediction for performance and power for CPU architectures. These works are orthogonal to ours, but illustrate the viability of the approach proposed herein.

\section{STT-RAM Cache Retention Time Prediction (SCART)}

\begin{figure}[t]
	   \vspace{-7pt}
		\centering
		\includegraphics[width=0.75\linewidth]{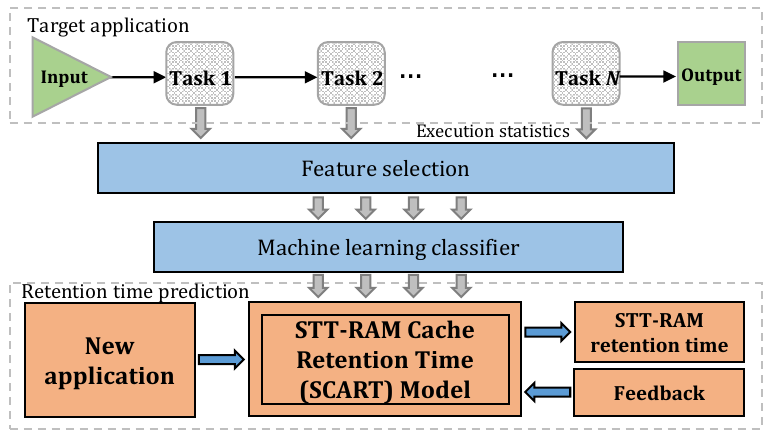}
		\vspace{-7pt}
		\caption{High-level overview of predictive model}
		\vspace{-15pt}
		\label{fig:model}
	\end{figure}

Unlike SRAM caches, where easily observable statistics from performance counters (e.g., cache miss rates) can be used to directly determine the best cache configurations, the correlations between miss rates and retention times are not that direct in STT-RAM caches. Therefore, in this work, we focus on using machine learning to predict the best retention times for STT-RAM L1 data cache energy and latency minimization based on hardware performance statistics. We chose to focus on the data cache since our experiments showed that the instruction cache blocks exhibit low variability in the retention time needs of the considered applications. A static retention time of $10ms$ sufficed for the applications considered. 


SCART incorporates a low-overhead machine learning classifier for design time or runtime fast and accurate prediction of retention times. For a design time exploration scenario, we assume that the target applications are first profiled on an SRAM cache with any arbitrary configurations. These statistics can be obtained via simulators (e.g., GEM5 \cite{gem5}) or by running the application on an actual computer. The execution statistics are then provided as input labels to SCART, which then outputs the best STT-RAM retention time for the target applications and specified objective function. This scenario is suitable for designing STT-RAM caches for an application-specific processor or provisioning a processor with a range of retention times in order to satisfy a variety of runtime retention time requirements \cite{Sun11,LARS8342053}. For a runtime scenario, the application can be run for a brief interval on one cache unit, and SCART uses the execution statistics to directly predict the best unit on which to run the rest of the application. SCART will substantially reduce the runtime complexity and migration costs for three system scenarios: 1) Multi-retention time cache designs (similar to \cite{LARS8342053}) for which the best cache unit must be determined during runtime; 2) hybrid caches to determine which unit to execute the application on; and 3) a multi-core system with a combination of SRAM and/or heterogeneous retention time STT-RAM caches \cite{arc_stt-ram}. 
\subsection{SCART Model Architecture} \label{sec:architecture}

Figure \ref{fig:model} presents a high level overview of our machine learning-based model. We model executing applications as task graphs, wherein each task may have one or more implementations, called \textit{task options} (e.g., different algorithmic implementations). These tasks are equivalent to application phases in our work. The different tasks and task options may have different execution characteristics, which also affect the target objective functions (energy or latency). Furthermore, each task may have different data configurations (e.g., data size, bit-width, etc.) that may change based on the inputs. 

The training data points are composed of execution statistics obtained from hardware performance counters. To generate the training data, we used GEM5 to gather the execution statistics of the different phases of a random subset of SPEC 2006, MiBench, and GAP benchmarks. We observed that 1 million instructions was sufficient to obtain stable statistics for predicting full phase behaviors. Thus, we used an interval size of 1 million instructions. As such, our model can predict retention times after executing an application or application phase for only 1 million instructions. 

Based on the SRAM characteristics of the training data, we performed feature selection to determine the most relevant features (i.e., hardware characteristics) for the STT-RAM retention time. We explored 59 features\footnote{The data can be found at \url{www.ece.arizona.edu/tosiron/downloads.php}} based on SRAM performance characteristics. These features can be either directly obtained from hardware performance counters or calculated from performance counter statistics. Some of the most important features included L1 and L2 cache miss rates, number of branches, cache read and write statistics while some less important features included the DRAM read and write bursts, number of integer and floating point instructions etc. 

To enable extensive testing, our initial training label size was 256 and the test label size was 64 (representing all the application phases). Our training label also consisted of six retention times: $10\mu s, 26.5\mu s, 50\mu s, 75\mu s, 100\mu s,$ and $1ms$. We empirically found that longer retention times were not beneficial for any of the considered applications. Given the selected features, we then fed the labels into a machine learning classifier (Section \ref{sec:classifier}) to develop SCART for predicting the best retention time for a new application.

To prevent substantial energy or latency degradation in runtime execution, the model also features a feedback mechanism that monitors the statistics of the predicted retention time. If the predicted retention time degrades the energy or latency compared to the base, the configuration is reverted to the base. To prevent data corruption resulting from the reduced retention time, we incorporate a low-overhead \textit{monitor counter}, similar to prior work \cite{LARS8342053,Sun11}, to keep track of each cache block's lifetime and invalidate the block (or write back to lower level memory if dirty) before the retention time expires. The counter can be implemented as an $N$-state finite state machine, which begins at the initial state when a block is written into the cache, counts up until the retention time is about to expire, and raises a flag to evict the block or write back to a lower memory level. We assumed $N$ = 4 in our work, resulting in a hardware overhead of only two bits per block.

\subsection{Machine Learning Classifier Comparison and Selection} \label{sec:classifier}

\begin{table*}[t]

\renewcommand{\arraystretch}{0.8}
\caption{Classifier F-scores}
\label{tab:classifers_comp}
\centering
\vspace{-1pt}
\scalebox{0.9}{
\begin{tabular}{|c|c|c|c|c|c|c|c|c|c|c|c|c}
    \hline
     & Linear SVC &Decision Tree &Extra Trees &Random Forest &KNN &RBF-SVC &Bagging &Adaboost &Gradient Boost\\  
    \hline
    F-score (0-1) &0.54713
&0.70989
&0.78098
&0.73437
&0.78203
&0.66875
&0.75625
&0.67552
&0.76692 \\

    
    \hline
\end{tabular}}
\vspace{-2pt}

\end{table*}

SCART features a machine learning classifier that comprises of two stages: the \textit{training stage} and the \textit{prediction stage}. In the training stage, the model learns the patterns in the input data (benchmarks and execution characteristics) and their correlations to the different retention time labels. In the prediction stage, the model takes as input new benchmarks and their characteristics, and outputs the predicted retention time labels for the new benchmarks. 

To select the best classifier, we considered several different classifiers and evaluated their accuracy. The classifiers we explored included: \textit{linear SVC, radial basis function SVC, decision tree, random forest classifiers, decision trees-based bagging, adaptive boosting, gradient adaptive boosting \cite{ml_algo_book}, extra-tree classifiers based ensemble technique \cite{extra_tree},} and \textit{K-nearest neighbor (KNN) classifiers \cite{knn}}. For brevity, we omit detailed descriptions of these classifiers, since they are described in prior work.

Table \ref{tab:classifers_comp} presents the different classifiers' F-scores \cite{f-measure}. The F-score is an evaluation metric that considers both precision and recall, and is a measure of a classifier's accuracy. The classifiers with the highest F-score were KNN and extra trees. However, we chose KNN classifier for use in our model due to its simplicity and lower prediction time (which makes it suitable for runtime predictions). Furthermore, KNN offers other advantages, such as lack of generalization (resulting in rapid training), and its non-parametric qualities. That is, KNN does not make any assumptions on the underlying data distribution. Thus, our model is amenable to applications that may not obey the typical theoretical assumptions (e.g., Gaussian mixtures, linearly separable, etc.). In general, KNN operates based on feature similarity; it determines how closely out-of-sample features resemble a training set, and classifies a given data point based on the similarity. Additional low level details of the KNN classifier can be found in \cite{knn}.
\subsection{KNN Classifier Tuning}

We observed that predicting the best retention times for latency vs. energy required different sets of features and KNN classifier characteristics. This observation was due to the conflicting nature of latency and energy with respect to retention time requirements. Thus, we tuned the KNN classifier and number of features to enable high accuracy for predicting retention times for latency or energy settings. Furthermore, to ascertain the robustness of our model, we randomly shuffled the data and performed five-fold cross-validation to ensure the validity of the classifier for a wide variety of applications. 

We empirically determined that the KNN classifier with three nearest neighbors and uniform weights achieved the highest F-score for latency and energy optimization. To select the appropriate features, we determined the features' importance values (that is, the features' impacts on prediction accuracy) using the \textit{scikit-learn} tools \cite{scikit-learn}, ordered the features in order of their importance, and eliminated the least important features for both latency and energy. 

\begin{figure}[t]
\centering
    \begin{subfigure}[t]{\linewidth}
      \centering
      \includegraphics[width=0.8\linewidth]{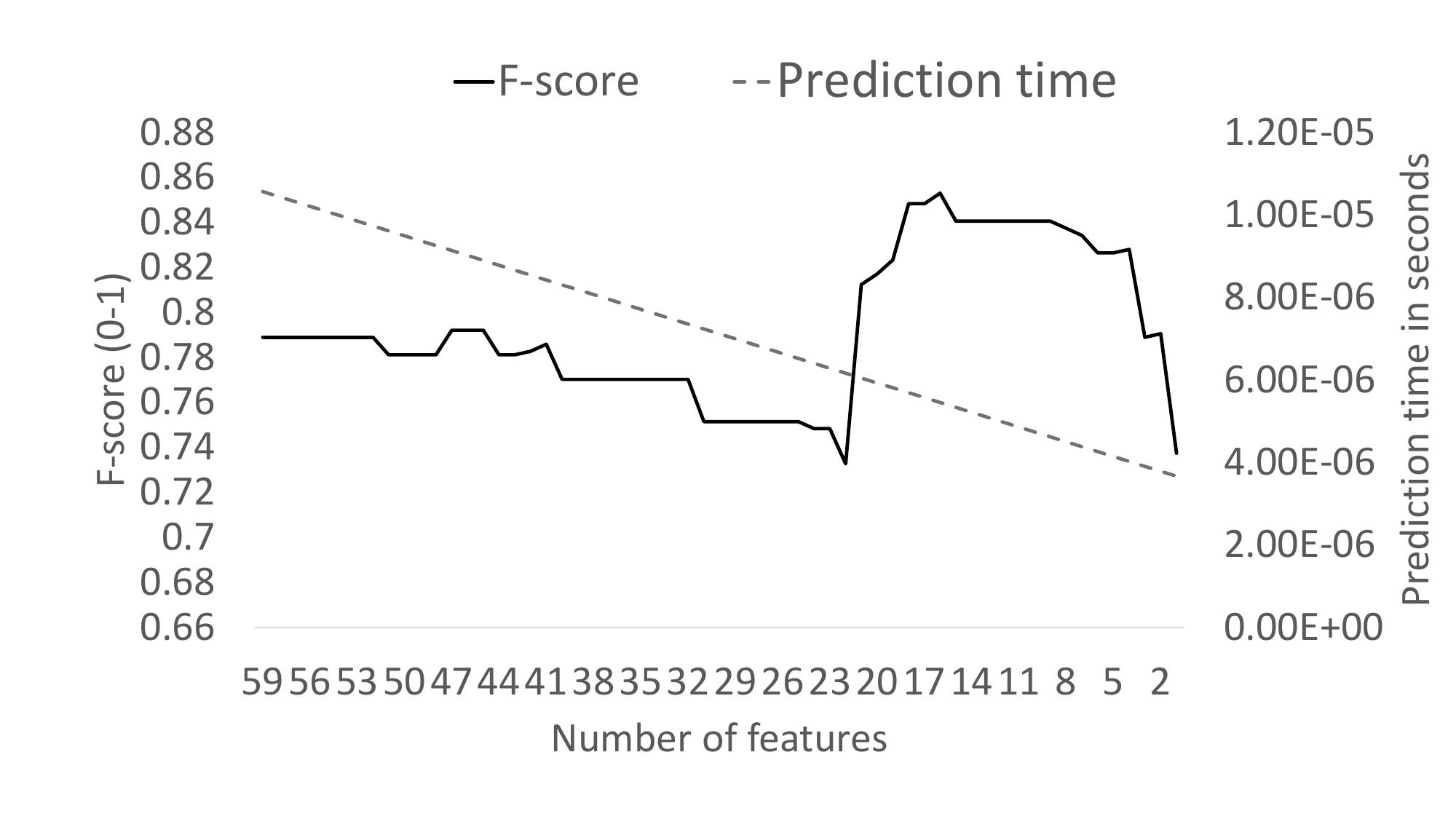}
      \vspace{-3pt}
      \caption{Latency}
      \label{fig:latency_fs}
    \end{subfigure}%

    \begin{subfigure}[t]{\linewidth}
      \centering
      \includegraphics[width=0.8\linewidth]{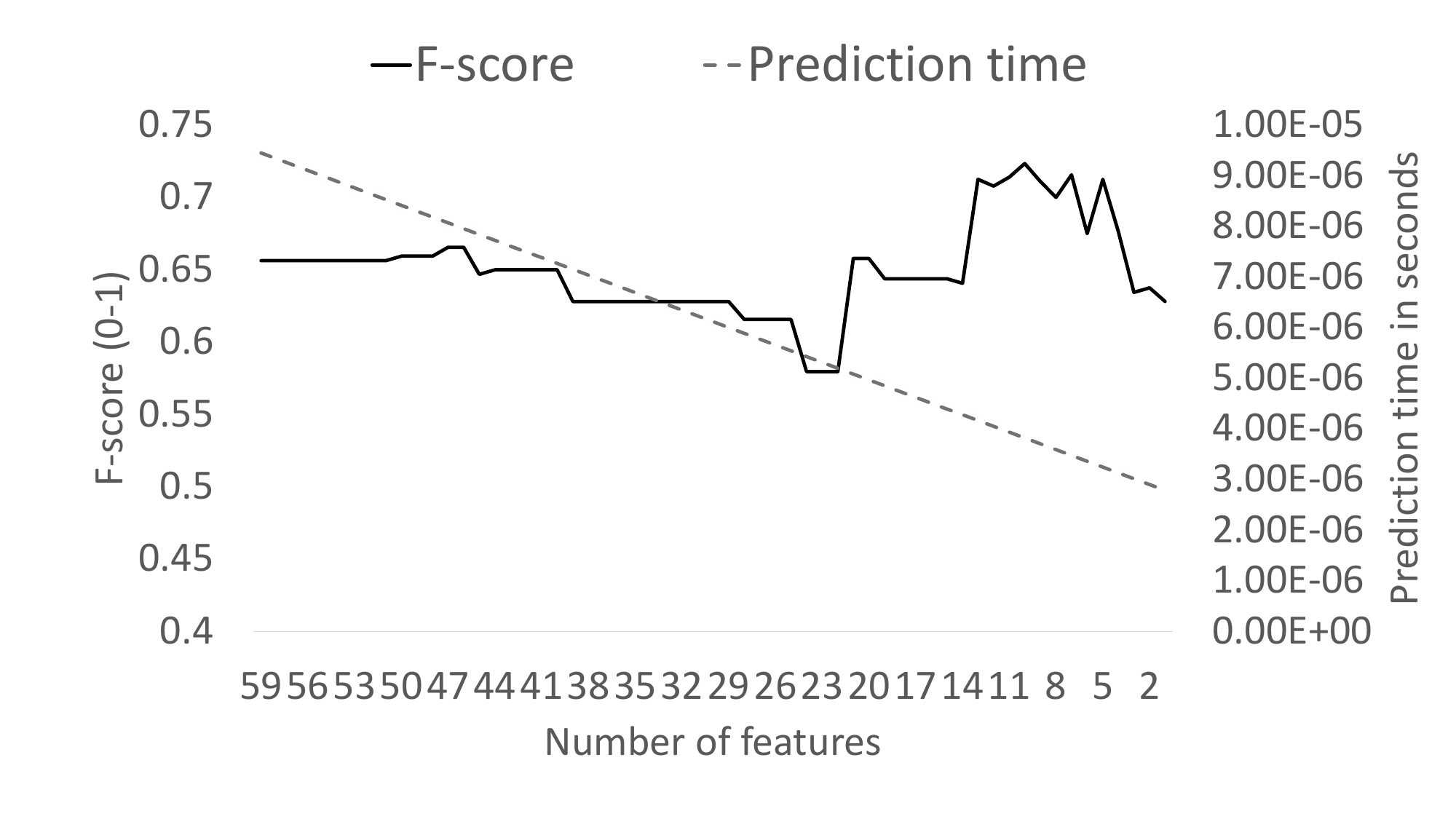}
      \vspace{-3pt}
      \caption{Energy}
      \vspace{-5pt}
      \label{fig:energy_fs}
    \end{subfigure}
    
\caption{Selection of optimal number of features for latency and energy optimization. Tuning began with 59 features, and features were iteratively removed to maximize F-score and minimize prediction time.}
\label{fig:feature_selection}
\vspace{-15pt}
\end{figure}

Figure \ref{fig:feature_selection} illustrates our selection of the optimal number of features for energy and latency. We compared the F-score and prediction time while iteratively eliminating the least important or redundant features in every run. The goal of iteratively eliminating the least important features was to find the optimal number of features that enabled the classifier to achieve the highest F-score. That is, we selected the fewest number of features, while eliminating features that did not change the F-score, since fewer features also reduce the prediction time.

    \begin{table*}[ht]
    \vspace{-5pt}
    \renewcommand{\arraystretch}{0.65}
    \caption{SRAM and STT-RAM cache parameters}
    \label{tab:retention}
    \centering
    \vspace{-6pt}
    \begin{tabular}{c||c|cccccc}
        \hline
        L1 cache configuration				&\multicolumn{7}{c}{32KB, 64B line size, 4-way}\\
        \hline
        L2 cache configuration				&\multicolumn{7}{c}{1MB SRAM, 64B line size, 16-way}\\
        \hline
        Memory device				&SRAM	    &\multicolumn{6}{c}{STT-RAM} \\
        \hline
        Retention times &--   &10$\mu$s     &26.5$\mu$s   &50$\mu$s   &75$\mu$s   &100$\mu$s	&1ms\\
        \hline
        Hit latency     	&0.486ns	        &0.464ns    &0.454ns 			        &0.448ns		        &0.445ns 		        &0.443ns    &0.438ns\\
        \hline
        Write latency    	&0.350ns	        &0.601ns    &0.769ns			        &0.894ns		        &0.981ns		        &1.045ns    &1.647ns \\
        \hline
        Read energy (per access)	&0.0076nJ    &0.003nJ   &0.003nJ	        &0.003nJ        &0.003nJ        &0.003nJ    &0.003nJ\\
        \hline
        Write energy (per access) 	&0.0066nJ    &0.026nJ   &0.030nJ	        &0.033nJ        &0.035nJ        &0.036nJ    &0.051nJ\\
        \hline
        Leakage power               &34.265mW     &\multicolumn{6}{c}{4.659mW}		\\
    
        \hline
    \end{tabular}
    \vspace{-10pt}
    
    \end{table*}

From Figure \ref{fig:latency_fs}, we observe that the highest F-score for latency optimization was obtained using 9 to 15 features. Thus, we used 9 features for latency in order to achieve a fast prediction time. For energy, as depicted in Figure \ref{fig:energy_fs}, 10 features achieved the highest F-score. Therefore, for both latency and energy, we eliminated the least important features until 9 and 10 features, respectively, remained. We also observed from our experiments that even though the highest accuracy was approximately 75\%, the false predictions still resulted in near-optimal retention times. As a result, SCART was able to achieve substantial latency and energy savings despite the error rate (Section \ref{sec:exhaustive}).

\section{Experimental Setup}

We modified the GEM5 simulator \cite{gem5} to model accurate STT-RAM behavior for different retention times and to capture the L1 and L2 cache statistics. We simulated single and quad-core processors with configurations similar to the ARM Cortex A-15 processor, with a 2GHz clock frequency. Each core had private instruction and data STT-RAM L1 caches, and a shared SRAM L2 cache (in the quad-core processor). Table \ref{tab:retention} depicts the cache parameters for both the SRAM and STT-RAM caches.

To represent a variety of workloads, we used 34 benchmarks in total (for both training and testing---see Section \ref{sec:architecture}); 22 from SPEC CPU2006 \cite{spec2006} (high performance benchmarks), 6 from MiBench \cite{mibench} (embedded systems benchmarks) and 6 from GAP \cite{beamer2015gap} (graph algorithms). We ran simulations for a maximum of one billion instructions for all the benchmarks, using the \textit{reference} and \textit{large} input sets for SPEC and MiBench, respectively, and 2048 nodes for the GAP benchmarks. We used Simpoint \cite{simpoint} to obtain the program phases for all the benchmarks, with intervals of 1 million instructions. We used execution statistics gathered after 1 million instructions for prediction.

For a thorough analysis, we initially considered nine retention times: 10$\mu$s, 26.5$\mu$s, 50$\mu$s, 75$\mu$s, 100$\mu$s, 1$m$s, 10$m$s, 100$m$s, and 1s. However, we found that the best latency or energy retention times for different applications were, for the most part, in the range of 10$\mu$s to 1ms. Thus, we eliminated 10$m$s to 1s from our modeling and analysis. To model the different retention times, we used the MTJ modeling technique proposed in \cite{Chun13} to compute the write pulse, write current and MTJ resistance value R$_{AP}$. We then applied the values to NVSim \cite{NVSim} and integrated with statistics  obtained from GEM5 \cite{gem5} to calculate the cache latency and energy. To model the SRAM cache in the hybrid cache, we used NVSim's SRAM settings. Table \ref{tab:retention} shows different latency and energy specifications for SRAM and STT-RAM used in our experiments. For stringent comparison, we used a hit cycle of 1 for both SRAM and STT-RAM, unlike prior work that used higher hit cycles for SRAM (e.g., \cite{Sun11}), thus resulting in lower optimization compared to SRAMs. To implement the machine learning algorithms, we used Python's \textit{scikit learn (Sklearn)} library \cite{scikit-learn}.

\section{results}

In this section, we first evaluate SCART in the context of a single-core processor, in comparison to a base retention time and exhaustive search. Thereafter, we evaluate SCART in the context of a quad-core processor running multi-programmed workloads, and finally compare SCART to prior work.

\subsection{Comparison to the Base Retention Time}

\begin{figure*}[t]
	    \vspace{-7pt}
		\centering
		\includegraphics[width=0.6\linewidth]{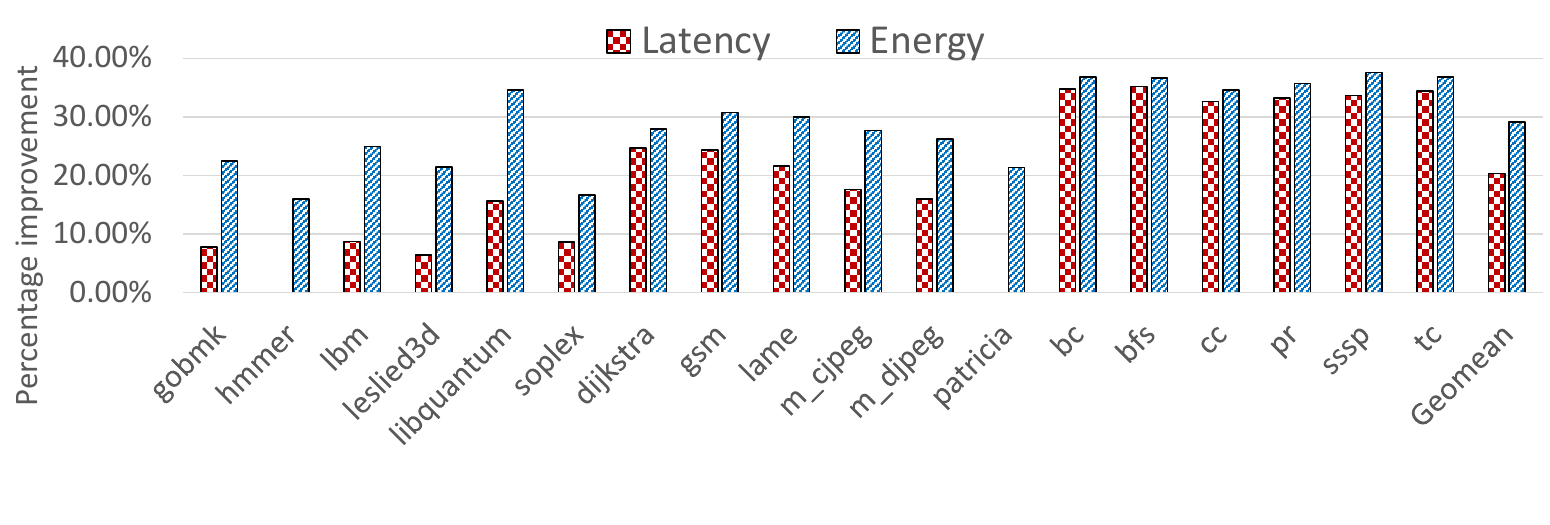}
		\vspace{-20pt}
		\caption{Percentage latency and energy improvements using SCART model compared to the base retention time of 1ms.}
		\vspace{-10pt}
		\label{fig:bench_suites}
	\end{figure*}

To evaluate SCART's effectiveness, we compared the latency and energy savings achieved by our model with a base retention time. We selected the base retention time as 1$m$s to be conservatively large enough to satisfy the cache block lifetimes of the considered applications, in order to prevent the need to refresh any blocks. Thus, the base configuration eliminates the additional overheads from refreshing data blocks \cite{Sun11}. For each benchmark, we report the overall results as the weighted combination of the phase results, as is the common practice in phase-based optimization \cite{simpoint}. 

Figure \ref{fig:bench_suites} depicts the latency and energy improvements achieved using SCART as compared to the base. On average across all the benchmarks, SCART improved the latency by 20.34\%, with improvements of up to 35.19\% for $bfs$ (breadth-first search algorithm). We observed different trends for different benchmark suites. For instance, SCART achieved substantial improvements over the base for the GAP benchmarks, since the base retention time was over-provisioned for the benchmarks. Most of the cache blocks needed to remain in the cache for much less than 1$m$s. On the other hand, SCART did not achieve substantial latency improvements for some SPEC and MiBench benchmarks, such as $patricia$, for which there was no improvement, and $hmmer$, for which SCART reverted to the base retention time in order to prevent a latency degradation. For $patricia$, the base 1$m$s retention time was sufficient for its cache block lifetimes, while $hmmer$'s cache blocks required more than 1$m$s retention time to prevent premature eviction. A closer look at $hmmer$'s cache blocks revealed that while several of the blocks required less than 1$m$s, there were also several blocks that required closer to 10$m$s to prevent premature expiry. However, using a 10$m$s base retention time would have incurred overall overheads for our mix of benchmarks. 

Similar to latency, SCART improved the energy, compared to the base, by an average of 29.12\%, with savings of up to 34.54\% for $libquantum$. The energy trends varied for the different benchmark suites, and we also observed that the retention time that was best for energy was not necessarily best for latency. For example, when SCART was set to optimize for energy, there was a latency overhead of 15.45\%; when it was set to optimize for latency, there was an energy overhead of 10.81\%. For a few benchmarks (e.g., $hmmer$), however, similar retention times sufficed for both latency and energy optimization. In general, SCART was able to trade off the non-optimized metric for the specified metric, as necessary. 

\vspace{-7pt}
\subsection{Comparison to Exhaustive Search} \label{sec:exhaustive}

To further evaluate SCART, we compared the results obtained to exhaustive search of the retention time design space. Note that while the retention time design space will typically not be expansive (six options, in our case), the design time overhead from exhaustive search comes into play when several applications or application domains must be explored. Thus, SCART must be able to rapidly determine retention times that are close to the optimal. 

\begin{figure}[t]
	    \vspace{-7pt}
		\centering
		\includegraphics[width=0.85\linewidth]{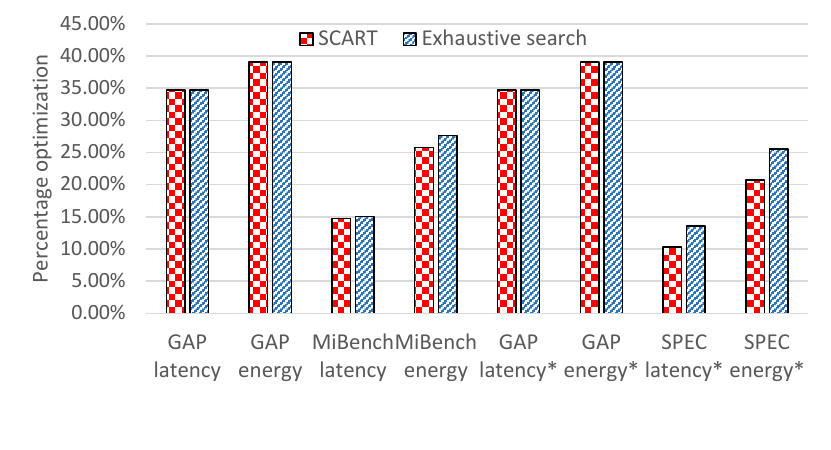}
		\vspace{-20pt}
		\caption{SCART vs exhaustive search latency and energy improvements compared to the base (1$m$s) retention time. Geometric means of the results are presented.}
		\vspace{-15pt}
		\label{fig:scart_comp}
	\end{figure}

Figure \ref{fig:scart_comp} depicts the comparison of the latency and energy achieved by SCART and exhaustive search (i.e., optimal) to the base. For brevity, we only show the geometric means for each benchmark suite considered. As seen in the figure, SCART's results were very close to exhaustive search for the different benchmark suites. For the GAP benchmarks, using SPEC benchmarks as training data, SCART achieved identical savings to exhaustive search for latency, and achieved energy savings within 0.07\% of the optimal. Similary, using SPEC benchmarks as training data for the MiBench workloads, SCART achieved latency and energy savings that were 0.4\% and 1.9\%, respectively, \textit{less} than exhaustive search. The degradation with respect to exhaustive search resulted from false prediction penalty of the labels. However, the penalty was low, since SCART predicted retention times that were close to the optimal, further illustrating SCART's effectiveness.

To further evaluate SCART's robustness, we also performed experiments to predict the retention times for GAP and SPEC benchmarks using training data from MiBench benchmarks (MiBench $\rightarrow$ GAP). We indicate the summary of the results for MiBench $\rightarrow$ GAP and MiBench $\rightarrow$ SPEC predictions in Figure \ref{fig:scart_comp} with an asterix (*). SCART achieved similar results to exhaustive search for MiBench $\rightarrow$ GAP predictions with average latency and energy savings of 34.71\% and 39.11\% over the base. However, while MiBench $\rightarrow$ SPEC yielded average latency and energy improvements of 10.3\% and 20.71\%, respectively, these results were farther from the optimal by 3.26\% and 4.87\%, respectively. We attribute this to the fact that the SPEC benchmarks' labels featured much higher variation than MiBench. As a result, a MiBench $\rightarrow$ SPEC prediction afforded less coverage in predicted characteristics than the SPEC $\rightarrow$ MiBench prediction. Overall, these results further illustrate SCART's ability to effectively predict latency and energy-saving retention times. 

\vspace{-5pt}

\subsection{SCART Execution in a Multi-Programmed Scenario}

\begin{table*}[t]

\renewcommand{\arraystretch}{0.65}
\caption{Multi-programmed workload distribution}
\label{tab:multi_label}
\centering
\vspace{-1pt}
\scalebox{0.75}{
\begin{tabular}{|c|c|c|c|c|c|c|c|c|c|c|c|c|c}
    \hline
    \# & Workload1 &Workload2 &Workload3 &Workload4 &Workload5 &Workload6 &Workload7 &Workload8 &Workload9 &Workload10 &Workload11 &Workload12\\  
    \hline
    1 & bc\_20	&dijkstra &m\_djpeg &cc\_20 &pr\_20 &gsm &tc\_20 &m\_cjpeg &patricia &bfs &sssp\_20 &lame\\
    2 &patricia & sssp\_20 &lame &gsm &sssp\_20 &pr\_20 &bc\_20 &bfs\_20 &m\_cjpeg &tc\_20 &m\_djpeg &dijkstra\\
    3 &gsm &sssp\_20 &tc\_20 &bc\_20 &pr\_20 &cc\_20 &patricia &bfs\_20 &lame &m\_cjpeg &m\_djpeg &dijkstra\\
    4 &sssp\_20 &gsm &tc\_20 &dijkstra &patricia &pr\_20 &m\_cjpeg &lame &bc\_20 &cc\_20 &bfs\_20 &m\_cjpeg\\
    \hline
\end{tabular}}
\vspace{-10pt}

\end{table*}

\begin{figure}[t]
	    \vspace{-1pt}
		\centering
		\includegraphics[width=0.85\linewidth]{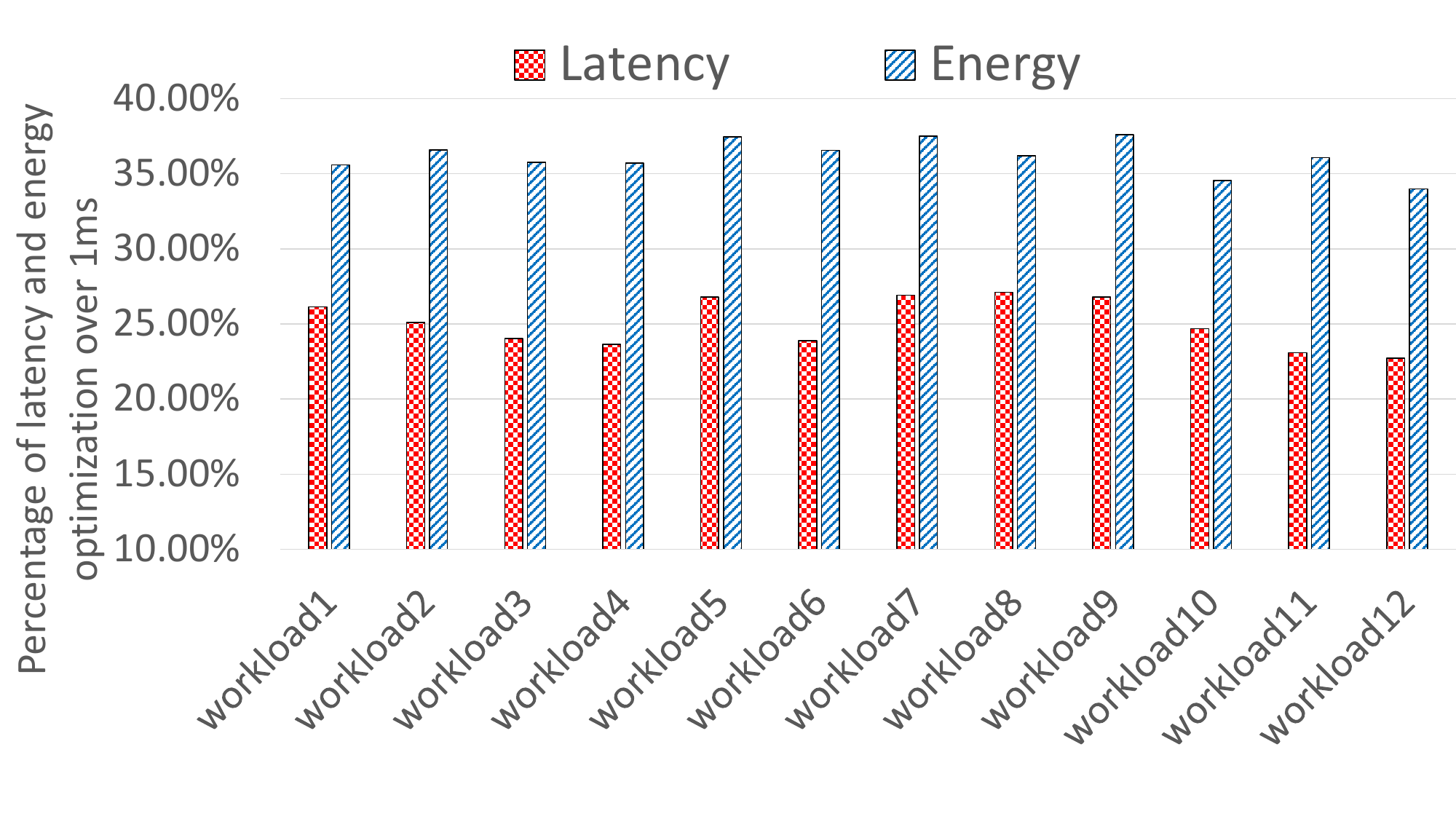}
		\vspace{-7pt}
		\caption{SCART latency and energy savings in a multi-programmed scenario}
		\vspace{-15pt}
		\label{fig:scart_workload}
	\end{figure}

To further evaluate our model, we tested SCART in a multi-programmed execution scenario featuring a quad-core processor with a shared 1MB L2 cache. The experiments performed herein enable us to evaluate SCART's scalability in a more complex system, since resource sharing in the L2 cache can impact the L1 cache behavior of the applications running on each core \cite{impact_resource_sharing}. We assume that each core features multiple retention time units as in \cite{LARS8342053}, and SCART predicts the best retention time unit for each application on each core. 

For the multi-programmed workloads, we created twelve workloads featuring a random combination of four benchmarks per workload, wherein each core runs one benchmark. The workloads used are shown in Table \ref{tab:multi_label}.  For the experiments in this subsection, we used the SPEC benchmarks (66\% of the total benchmarks) as training data and MiBench and GAP benchmarks (33\%) as testing data.

Figure \ref{fig:scart_workload} summarizes the percentage latency and energy optimizations achieved by SCART in the multi-programmed scenario compared to a base retention time of 1$ms$. On average across all the workloads, SCART achieved latency and energy savings of 25.07\% and 36.13\%, respectively. As seen in Figure \ref{fig:scart_workload}, the latency and energy savings were relatively consistent across the different workloads, demonstrating SCART's effectiveness in various execution scenarios.
\vspace{-5pt}
\subsection{Comparison to Prior Work and Implementation Overhead}

To further evaluate the effectiveness of our approach, we compared the exploration time to prior work \cite{LARS8342053} that proposed different retention time units within each STT-RAM cache. We chose this prior work, called \textit{LARS}, since it is the most related to ours and determined the optimal latency and energy configurations during runtime using exhaustive sampling. However, unlike LARS, which had four retention times, our implementation featured six retention times. In our implementation, each benchmark was first run on the base STT-RAM unit (1$m$s) for 1 million instructions, and the data was then used by SCART to predict the best retention time unit on which to run the rest of the application. Overall, SCART achieved similar results to exhaustive search (Section \ref{sec:exhaustive}). 

Given SCART's similar performance to exhaustive search, we also evaluated SCART's benefit for reducing the exploration/tuning time. In LARS, the applications were sampled on each STT-RAM cache unit. Thus, LARS required six migrations between cache units for each tuning decision, with each migration taking 4608 cycles, which translates to 2.304$\mu$s at a 2GHz frequency. In total, the migration overhead was 13.824$\mu$s. SCART, for most of the cases, required only one migration if a different retention time than the base was determined to be the best. Therefore, SCART's average overhead (prediction + migration) was 6.554$\mu$s, reducing the exploration overhead by 52.58\% compared to LARS, while achieving similar latency and energy savings. Furthermore, unlike LARS, which runs the application on potentially sub-optimal retention times before arriving at the best, SCART directly predicts the best without exploring sub-optimal retention times. 

We assume that SCART is implemented in software (e.g., in the operating system). As such, SCART does not incur any hardware overhead other than the monitor counter described in Section \ref{sec:architecture}. However, SCART incurs some memory overhead. We used memory profiling to observe the memory consumed by SCART, and found that SCART consumes 0.156 MB of memory during the training stage and 2.5 KB of memory for the runtime prediction stage.
\section{Conclusion and Future Work}
In this paper we proposed an STT-RAM Cache Retention Time (SCART) model that uses a KNN classifier to predict the best retention time for an STT-RAM L1 cache. SCART uses execution statistics obtained from hardware performance counters. In a runtime single-core scenario, SCART predicted retention times that achieved average latency and energy savings of 20.34\% and 29.12\%, respectively, compared to a base 1$m$s retention time. In a quad-core scenario with multi-programmed workloads, SCART achieved average latency and energy savings of 25.07\% and 36.13\%, respectively, compared to a base 1$m$s retention time. Compared to prior work, SCART reduced the exploration time by 52.58\%, while achieving similar latency and energy savings. Future work involves exploring a hardware implementation of SCART, extending SCART to predict other architecture parameters, and reducing the number of required labels in order to reduce the memory overhead, without sacrificing prediction accuracy. 

\section*{Acknowledgement}
This work was supported in part by the National Science Foundation under grant CNS-1844952. Any opinions, findings, and conclusions or recommendations expressed in this material are those of the authors and do not necessarily reflect the views of the National Science Foundation.
	\renewcommand{\bibfont}{\small}
	\balance
	\bibliographystyle{IEEEtran}
	\bibliography{refs}
	
\end{document}